\documentclass{IEEEtran}
\usepackage{cite}
\usepackage{amsmath,amssymb,amsfonts}
\usepackage{graphicx}
\usepackage{comment}
\usepackage{textcomp,nicefrac}
\usepackage[T1]{fontenc}
\def\BibTeX{{\rm B\kern-.05em{\sc i\kern-.025em b}\kern-.08em
T\kern-.1667em\lower.7ex\hbox{E}\kern-.125emX}}
\begin{document}
\title{FERS-5200: a distributed Front-End Readout System for multidetector arrays}
\author{M.Venaruzzo, A. Abba, C. Tintori, and Y. Venturini
\thanks{Paper submitted for review on October 28$^{th}$ 2020.}
\thanks{M. Venaruzzo, Marketing Technical Assistance and Asia Area Sales Manager at CAEN SpA, Italy (e-mail: m.venaruzzo@caen.it).}
\thanks{A. Abba, Co-Founder at Nuclear Instruments srl, Italy (e-mail: abba@nuclearinstruments.eu).}
\thanks{C. Tintori, CTO at CAEN SpA, Italy (e-mail: c.tintori@caen.it).}
\thanks{Y. Venturini, Marketing Technical Assistance and Sales Manager at CAEN SpA, Italy (e-mail: y.venturini@caen.it).}}

\maketitle

\begin{abstract}
The FERS-5200 is the new CAEN Front-End Readout System for large detector arrays. It consists in a compact, distributed and easy-deployable
solution integrating front-end based on ASICs, A/D conversion, data processing, synchronization and readout. Using the appropriate Front-End the solution perfectly fits a wide range of detectors such as SiPMs, multianode PMTs, GEMs, Silicon Strip detectors, Wire Chambers, Gas Tubes, etc. The first member of the FERS family is the unit A5202, a 64 channel readout card for SiPMs, based on the CITIROC ASIC by Weeroc SaS. The Concentrator board DT5215 can manage the readout of up to 128 cards at once, that is 8192 readout channels in case of the A5202.
\end{abstract}


\section{Introduction}
\label{sec:introduction}
\IEEEPARstart{I}{n} the traditional approach to readout systems for physics, the detectors are usually interfaced to close-by Front-End Preamplifiers and long cables bring analog signals to the readout electronics (ADC, TDC, etc.), with A/D conversion, online data processing and communication interfaces concentrated in racks. Given the recent developments in the field of detectors technology, physics experiments are moving in the direction of using large arrays of detectors, to be read out with more compact and more cost-effective electronics. In particular, detectors such as SiPMs, MA-PMTs, GEMs, Silicon Strips, Gas Tubes, etc., are becoming more and more widely used to build huge experimental setups. CAEN, in collaboration with Nuclear Instruments Srl, has developed a new platform, called FERS-5200, to fit the requirements coming from this type of experiments.\\FERS-5200 is a \textbf{distributed} and \textbf{easy-scalable} system, where each Front-End unit is a small card that can play different roles such as: 

\begin{itemize}
	\item a traditional analog spectrocopy chain housing 32 or 64 channels with preamplifier, shaper, discriminator, A/D converter
	\item a digital front-end like a Time-to-digital converter (TDC) or a trigger logic board
	\item a Switched capacitor array
\end{itemize}

\begin{figure}
	\begin{center}
		\includegraphics[width=3.5in]{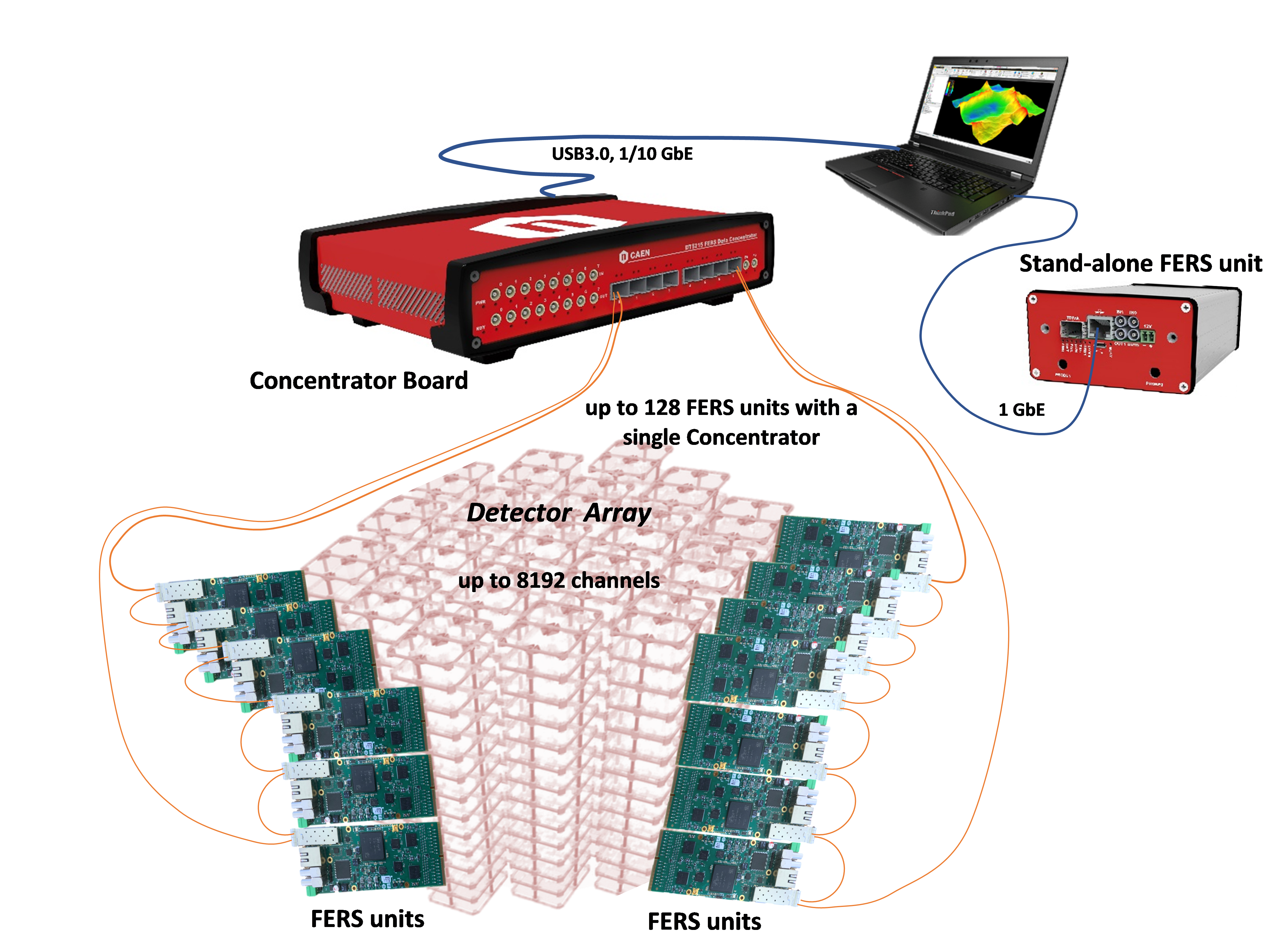}
	\end{center}
	\caption{FERS-5200 tree network: up to 16 FERS units can be daisy-chained via Optical Link (TDlink), the Concentrator Board is capable to read out 8 TDlink for a total of 128 cards. Communication interfaces to the host PC are standard Ethernet 1/10 Gb and USB 3.0. A sigle FERS units, housed in an aluminium box can be used as standalone, directly connected to the host PC via Ethernet or USB 2.0.}
	\label{fersnet}
\end{figure}

Each FERS card featurs also the synchronization capabilites, local memory and the readout interface. Multiple FERS units can be daisy-chained through a special protocol (TDlink) on optical fiber bringing slow control, readout and synchronization at once. Up to 16 FERS units can be daisy-chained and readout via a single link. Thanks to the Concentrator Board DT5215, hosting 8 TDLinks, it is possible to build a network (FERSnet) consisting of up to 128 FE cards and further extension of the system is possible by synchronizing more than one Concentrator Board (see Fig.\ref{fersnet}).\\ FERS is designed to be a \textbf{flexible} platform: keeping the same back-end (that is readout architecture and interface), different types of Front-End will be developed to fit a variety of detectors \cite{fers-5200}. Typically, the front-end is based on ASIC chips that allow for high channel density and cost-effective integration into small size and low power modules. The first developed unit is the A5202, that uses the CITIROC 1A chip produced by Weeroc SaS for SiPM readout \cite{citirocdatasheet}; in the next future there will be a complete line of FERS units using different ASICs or even preamps made of discrete components to match other type of detectors' requirements.

\section{Architecture of the Front-End}
\begin{figure}
	\begin{center}
		\includegraphics[width=3.5in]{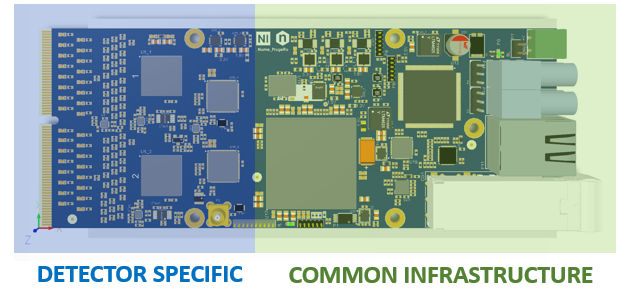}
	\end{center}
	\caption{General view of a FERS unit, where the detector specific part and the common infrastructure are highlighted.}
	\label{common}
\end{figure}

FERS-5200 is designed to be tailored for different specific detectors and applications. The Front-End is a compact card (nearly 15 x 6 cm) hosting the analog Front-End, ADC and/or TDC, FPGA, I/Os, communication interfaces and, in some cases, the detector bias power supply. In most cases the analog Front-End is based on ASICs, to optimize channel density and cost-effectiveness of the solution. \\ One FERS unit can be used stand alone, mainly for evaluation and basic applications, or in a network. In single mode,
the unit is directly connected to the computer via USB 2.0 or Ethernet 10/100T. For large readout systems, a flexible
and scalable network of units (FERSnet) can be created by means of the high speed optical link called TDlink, that
allows up to 16 FERS units to be connected in daisy chain (ring), providing data readout, synchronization between the
units and broadcasting of commands, such as triggers, time resets, etc.\\
The architecture of the FERS unit will be the common infrastructure to provide an easy integration of different analog Front-Ends, either ASICs, either discrete components (see Fig. \ref{common}). In this way a complete line of FERS units will be available, making FERS-5200 a platform covering the main types of experimental setups and applications in the field of nuclear and particle physics. \\
The first FERS units is the A5202 (or DT5202 in boxed version for stand-alone use) which is a 64-channel board for SiPM readout based on the CITIROC-1A ASIC by Weeroc SaS. 

\section{Easy-scaling using the TDlink}
\begin{figure}
	\begin{center}
		\includegraphics[width=3.5in]{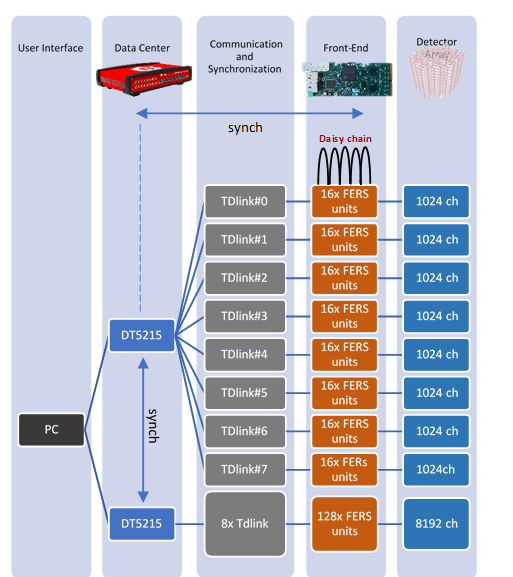}
	\end{center}
	\caption{The FERSnet: each Concentrator Board DT5215 is able to readout 16 FERS units per link, for a total of 128 cards. Multiple Concentrator Boards can be synchronized.}
	\label{schema}
\end{figure}

Typically, in large installations, the FERS unit is controlled and read out through the TDlink in a FERSnet using the Concentrator Board (see Fig. \ref{schema}). The TDlink is a timing and data link able to distribute a reference clock, broadcast synchronization and acquisition commands (start run, stop run, triggers) and read/write data packets for both readout and slow control. The physical layer of the TDlink is a 6.25 Gbit/s duplex link, running over optical fiber (LC connectors). The readout bandwidth of the TDlink is > 100 MB/s. When N FERS units are connected in daisy chain (N=1 to 16), the bandwidth is shared between the units.\\
The propagation delay of the data packets and commands circulating over the TDlink is not completely deterministic. In fact, the link performs occasional auto-synchronization procedures that require the transmission of low level data packets, not controlled by the user protocol. This packets have higher priority than the user payload and introduce an unpredictable delay. While this small uncertainty of the propagation delay is not an issue for the readout, it makes the link unsuitable for the distribution of timing signals and, sometimes, also for the distribution of the global trigger when low jitter and low propagation delay is mandatory. For these cases, the FERS unit provides additional LEMO connectors (2 inputs + 2 outputs) connected to the FPGA, implementing digital I/Os with low propagation delay.\\
The TDlink is able to guarantee that all the FERS units in the network are running with a synchronized global time: in fact, after the power on, the TDlink sends a series of synchronization packets that allow the link master to understand the number of connected slaves (FERS units) and the exact delay of each node. The synchronization procedure initializes the local time counter in each unit (set the same zero) and provides a low jitter clock signal (typ. 200 MHz) that is used for the internal timing of the boards.  The procedure takes into account the propagation delay within the network to compensate for the time skew. Eventually, all the channels in the system will be synchronized, meaning that all the 64 bit local times (absolute time) are locked and aligned. 

\begin{figure}
	\begin{center}
		\includegraphics[width=3.5in]{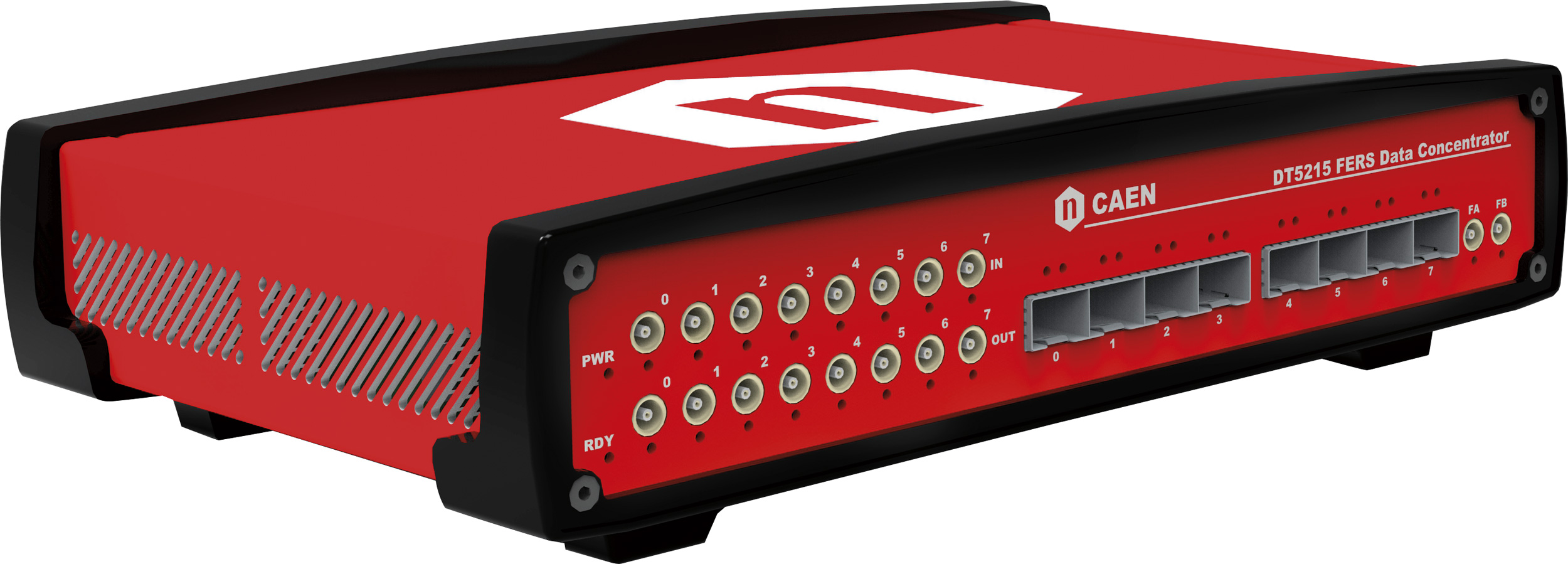}
	\end{center}
	\caption{General view of the Concentrator Board DT5215. The board is equipped with 8 TDlink for system building and a Xilinx Zynq Ultrascale SoC with embedded ARM. It is able to manage the readout process of up to 128 FERS units, perform event sorting, event building and data buffering.}
	\label{dt5215}
\end{figure}

The required bandwidth for each FERS unit depends on the acquisition mode. In Spectroscopy mode, the event data packet contains header (16 bit) + channel mask (64 bit) + trigger time stamp (48 bit) + one charge value per channel (16 bit). Without zero suppression, the event data size is 140 bytes. Considering 10$\mu$s conversion time, the maximum trigger rate is 100 KHz and the maximum throughput is 14 MB/s. It is realistic to have data reduction of ~80$\%$ after the zero suppression, thus reducing the throughput down to less than 3 MB/s/unit, which is below the available bandwidth also in the worst case of 16 units in daisy chain.\\
Besides the TDlink, the FERS unit provides also a USB 2.0 and an Ethernet interface (TCP-IP @ 10 or 100 Mb/s) that allow a single unit to be controlled and read out by a PC without any additional hardware. This solution is very convenient for a quick and easy startup and evaluation of a small system, but it is not ideal for large systems and for scalability because there is no synchronization between multiple units.\\
The FERS Concentrator Board (DT5215 shown in Fig.\ref{dt5215}) is a data collector housing eight TDlink masters, thus making possible to manage up to 128 FERS units, that is 4k-8k detectors, depending on the configuration of the Front-End card. The Concentrator Board is connected to the host computer through 1 or 10 Gbit Ethernet or USB 3.0. The total data throughput of the FERSnet is about 200 MB/s. The Concentrator has an embedded ARM processor (Quad Core) running Linux; it is therefore possible for the user to run a custom software on the board and process the data locally, before sending the event data to the computer for the analysis. The collector board provides a few GB of DDR memory for local data storage. More than one Concentrator modules can be synchronized by means of a dedicated daisy chainable S-link, where the first module acts as a master, generating the synch clock and the time stamp reset for the other modules in the chain. The synchronization with an external reference is possible too.\\

\section{A5202: the FERS unit for SiPM readout}
\subsection{Hardware description}
\begin{figure}
	\begin{center}
	\includegraphics[width=3.5in]{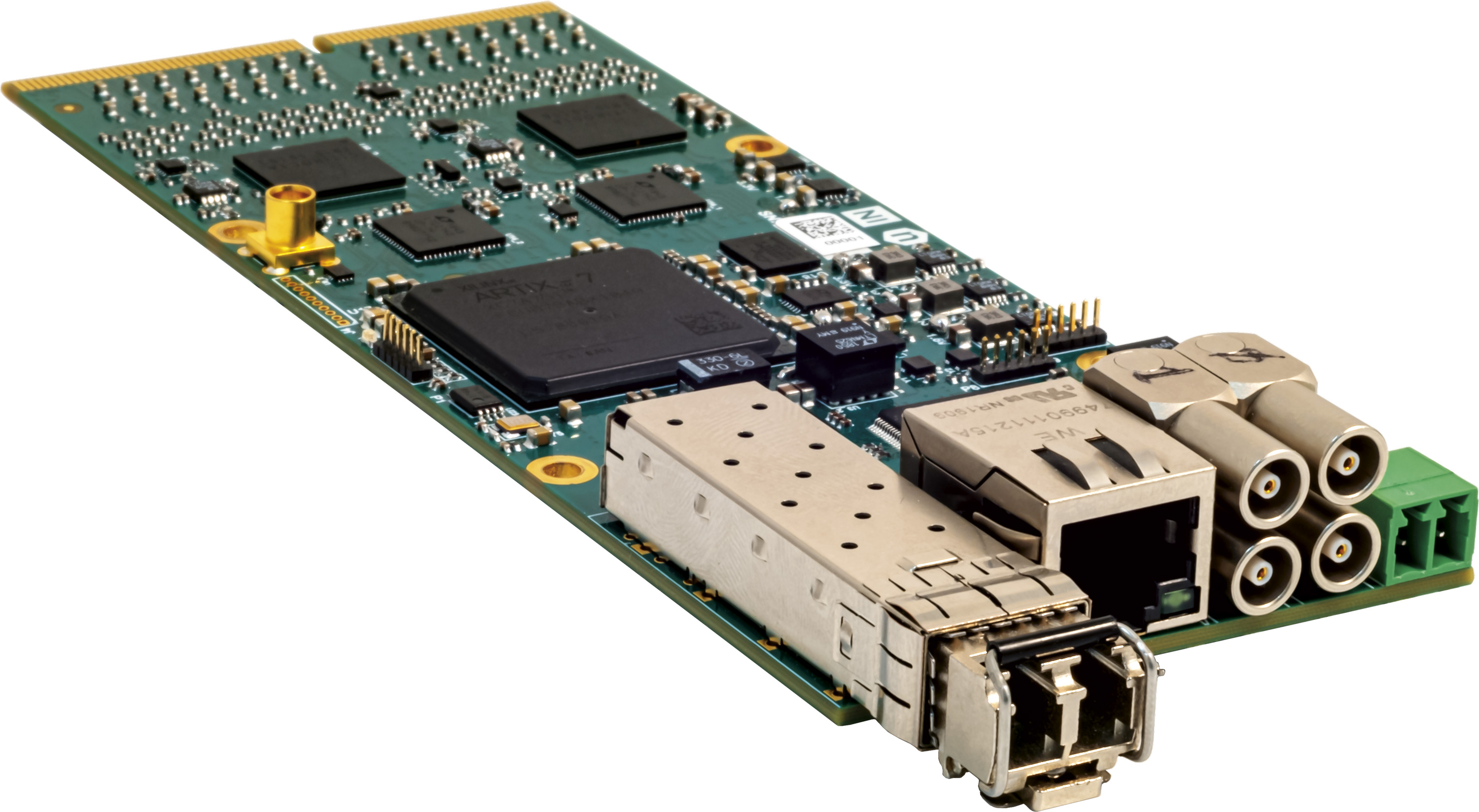}\quad\includegraphics[width=3.0in]{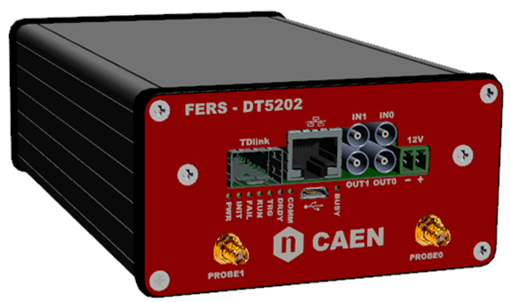}
	\end{center}
	\caption{General view of the 64-channel FERS unit for SiPM readout. Top: A5202 naked version. Bottom: DT5202 boxed version, particularly suited for evaluation stand-alone use.}
	\label{a5202}
\end{figure}

\begin{figure}
	\begin{center}
		\includegraphics[width=3.5in]{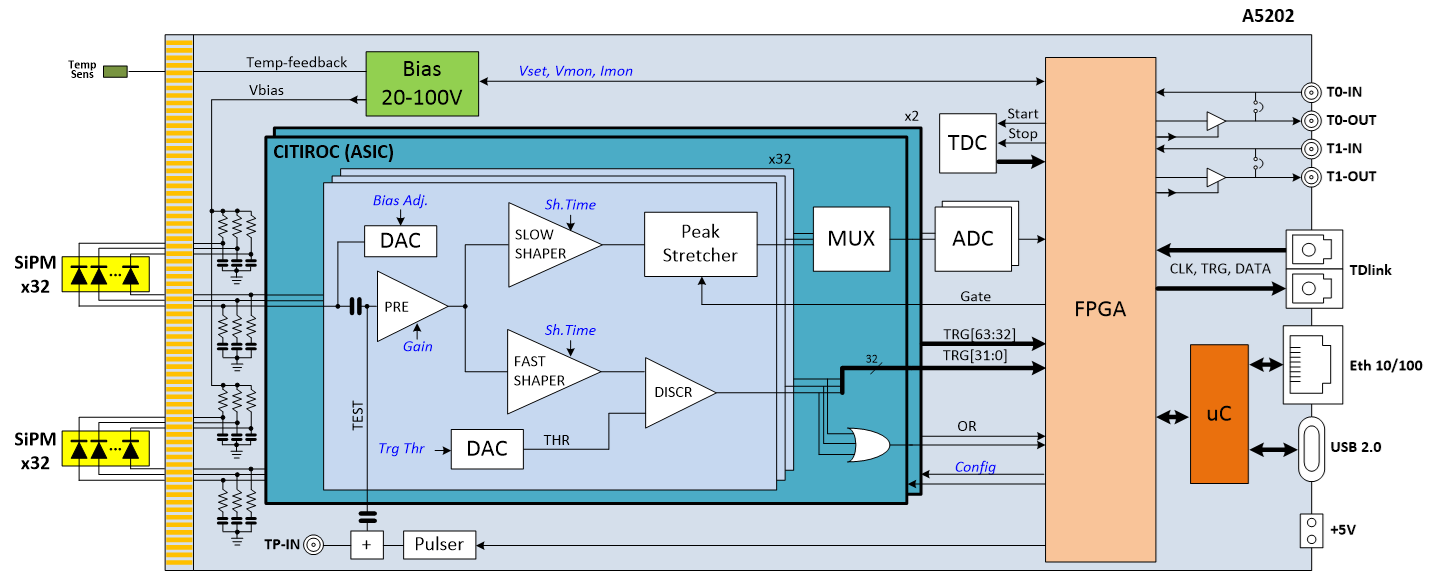}
	\end{center}
	\caption{Scheme of the A5202 FERS unit. Connections between the analog Front-End and the FPGA are shown.}
	\label{schema_a5202}
\end{figure}

The A5202 is a 64-channel unit for SiPM array or matrix readout. It is a small card ($\sim$ 6 x 15 cm) housing two CITIROC 1A ASICs, the ADCs, the FPGA, the bias power supply for the SiPMs and the interfaces for readout, synchronization and control (see Fig.\ref{a5202} and \ref{schema_a5202}).\\ 
CITIROC 1A is a 32-channel front-end ASIC designed to readout silicon photo-multipliers (SiPM) for scientific instrumentation applications (refer to \cite{citirocdatasheet}). CITIROC 1A allows triggering down to 1/3 photoelectrons (p.e.) and provides the charge measurement with good noise rejection and 1$\%$ linearity up to 2500 p.e. Moreover, it outputs the 32-channel triggers with a high resolution timing (better than 100 ps), although there is not an internal TAC/TDC to acquire the timing measurement. It is possible to use the FPGA of the A5202 for this purpose and implement a low resolution TDC (0.5 ns) to calculate the start-stop time between a reference signal and the input pulses. A better timing information can be achieved by the additional TDC (50 ps LSB) connected to the FPGA, providing the start-stop measurement between the OR of the self-triggers and asynchronous signal coming from T0 or T1 LEMOs (typically a reference timing signal or the global acquisition trigger). Refer to \cite{citirocdatasheet} for more details about the CITIROC features. \\

\begin{figure}
	\begin{center}
	\includegraphics[width=3.5in]{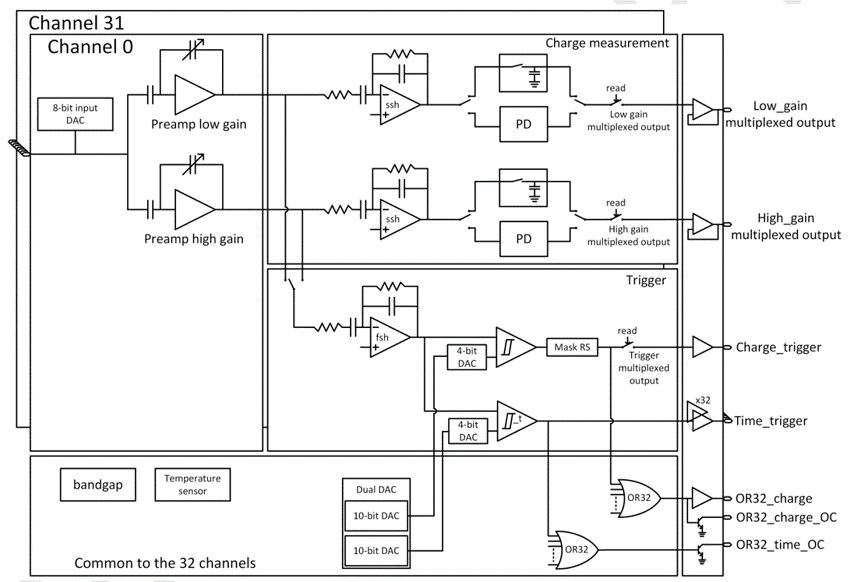}
	\end{center}
	\caption{Electrical scheme of the CITIROC 1A ASIC by Weeroc. Two of these chips are hosted on the A5202 card as analog Front-End for SiPMs.}
	\label{citiroc}
\end{figure}

\begin{figure}
	\begin{center}
	\includegraphics[width=2.5in]{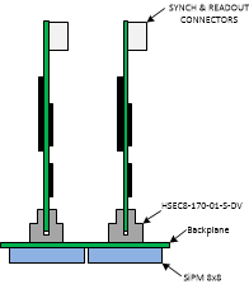}
	\end{center}
	\caption{Possible geometrical arrangement for the A5202 with backplanes for SiPM connection.}
	\label{backplane}
\end{figure}

Besides the CITIROCs, the A5202 embeds a programmable high voltage power supply (20-85V, 10 mA) for the bias of the SiPMs, featuring a feedback loop with the temperature sensor (internal or external) for the compensation of the gain drift. An individual fine adjustment of the high-voltage is possible using a channel-by-channel DAC connected to the ASIC inputs, thus allowing the correction of the non-uniformity of SiPMs. An internal calibration signal allows CITIROC 1A to be calibrated.\\
The A5202 has an input edge card connectors type HSEC8-170, mating to a Samtec HSEC8-170-01-S-DV. The connector has 140 contacts (0.8 mm pitch) and brings 64 couples (SiPM anode and cathode), the temperature sensor and several grounds. The edge connector makes it possible to build a backplane or a flange that houses the SiPMs detectors on one side and the housing for the A5202 on the other side (see Fig.\ref{backplane}). In this way maximum compactness of the system is achieved, together with reduced cable cost, small noise pickup, weak signal attenuation bulkiness and reduced risk of contact failures. CAEN provides also an adapter from the HSEC8-170 edge-connector to standard 2.54 mm pin headers.\\

\subsection{Acquisition modes}
The A5202 can work in different acquisition modes, according to the CITIROC capabilities. In particular, Photon Counting, Spectroscopy, Timestamping and Time-stamped Energy modes are implemented. These can be considered as possible operation modes for all types of FERS units: depending on the ASIC/Front-End architecture it will be possible to implement some of these readout modes or add some others (like, for instance, Waveform Recording mode).\\

\subsubsection{Photon Counting Mode (MCS)}
This mode is typically used for imaging, for instance with an 8x8 SiPM array. The trigger threshold for counting can be adjusted down to 1/3 p.e., thus allowing for single photon counting. The maximum counting rate is 20 Mcps.\\
The FPGA implements 64 counters connected to the individual self-triggers. The counters can be read on the fly (from the software) or simultaneously latched and stored into the local memory buffer with the trigger (MCS mode). In this mode, the counting intervals (or slots) are defined by the trigger that, in most cases, is an internal periodic signal with Dwell time in the range 1 $\mu$s to 10 s or an external trigger from LEMO T1. A data packet is produced for each interval: it contains a progressive index (slot ID), a 48 bit time stamp and 64 words with the value of the counters. The counters are reset after each interval. There is no dead time between two intervals.\\

\subsubsection{Spectroscopy Mode (PHA)}
The Energy mode is typically used for trackers, veto systems, calorimeters and also for spectroscopy applications. In the latter case, if PHA spectra are required, these must be created and populated in the software using the event list.\\
The Energy (or Charge) mode is always controlled by a trigger signal and the acquisition is simultaneous on the 64 channels of the unit. The trigger can be local (= OR or Majority of the 64 self-triggers) or, most likely, come from an external trigger logic that combines the self-triggers of the whole system (usually the combination of single FERS units' triggers made in the Concentrator).
When the trigger is asserted, the FPGA drives the CITIROC to hold the peak values detected in the shaper amplifiers, then starts the sequence for the A/D conversion. This takes about 10$\mu$s and causes a dead time for the acquisition, allowing for a maximum trigger rate of about 100 kHz.\\If enabled, after the A/D conversion the FPGA applies the Zero Suppression: channels that did not trigger the charge discriminator are not saved into the memory buffer.\\
Besides the absolute timestamp of the trigger given with a granularity of 8 ns, a fine time measurement coming from the high resolution TDC (50 ps) is also provided. It measures the time difference between the T-OR (=OR of the self-triggers generated by the timing discriminators) and an external time reference (feeding T0 or T1), for instance the RF signal of the accelerator or the bunch crossing.\\

\begin{figure}
	\begin{center}
	\includegraphics[width=7.0in,angle=90]{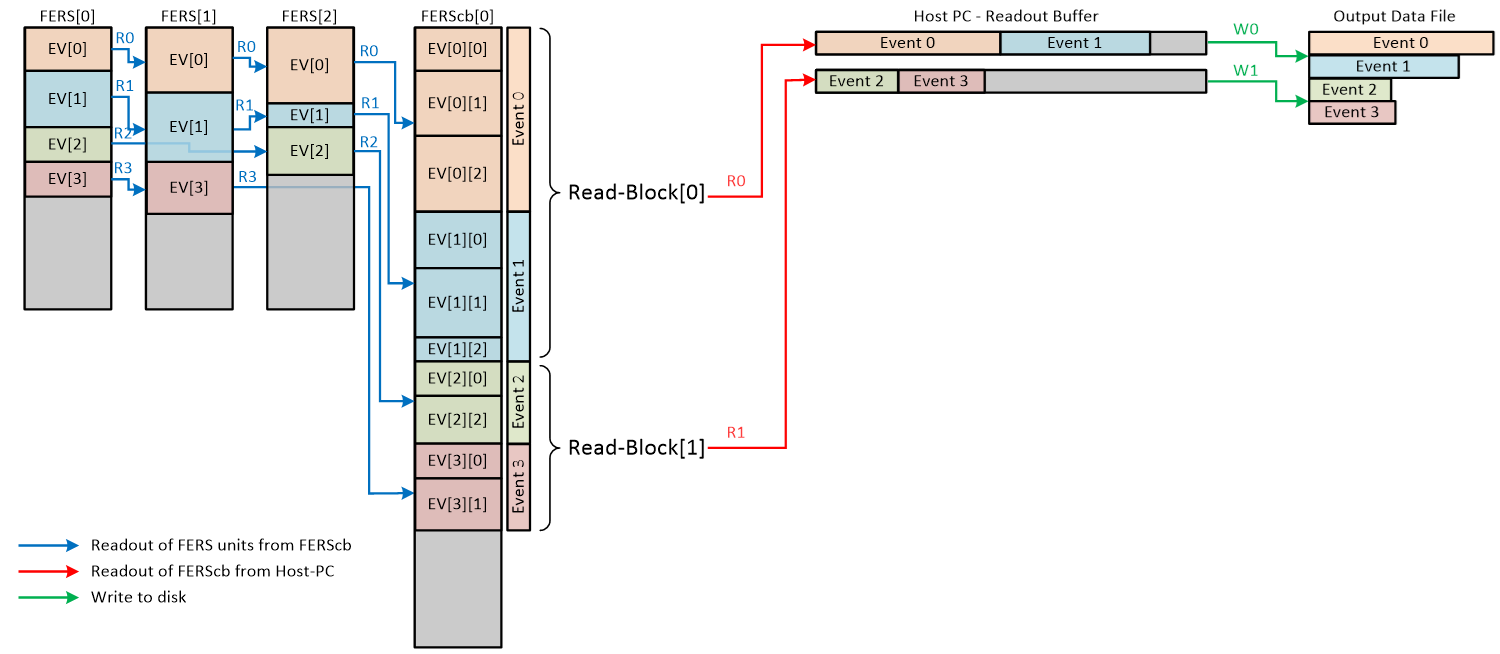}
	\end{center}
	\caption{Example of Horizontal Readout with 3 FERS units acquiring 4 events (with trigger loss on 1 and 2)}
	\label{horizontal_readout}
\end{figure}

The data readout in Energy mode is designed to facilitate the event building: data packets belonging to the same trigger and acquired by different FERS units are read out “horizontally”, meaning that each block read initiated by the DAQ software involves the packets of a specific trigger index (see Fig.\ref{horizontal_readout}). The first FERS unit in the system transfers the data packet of the trigger N, then passes a token to the next unit in the chain that appends its data packet, going on up to the last unit in the chain. Multiple chains are appended in the Concentrator modules. If one FERS unit missed a trigger for any reason, it will skip the readout of that trigger (does not provide any data), so that the packets alignment is always guaranteed.\\

\subsubsection{Timing Mode}
In Timing Mode, the channels acquire independently: for each self-trigger, the FPGA saves the channel ID (0 to 63) and the time stamp of the hit into the local memory buffer. The output data is therefore a list of time-stamped events grouped in packets of programmable size. The time stamp can be referred to the start of the acquisition (absolute time) or obtained as a start-stop measurement (delta-T) with respect to a reference signal coming from the input T0. In both cases, the time stamp has a maximum resolution of 0.5 ns and is expressed with 24 bits, so the dynamic range is ~8.4 ms. It is possible to program the LSB in order to have less resolution and more dynamic range. \\

The Timing mode supports 3 sub-mode:

\begin{itemize}
	\item Common start: only the events falling within a defined time window opened by a reference signal are acquired; 
	\item Common Stop: only the events falling within a defined time window closed by a reference signal are acquired;
	\item Streaming: all the events are acquired.
\end{itemize}

The time stamping mode is dead time free as far as the throughput does not saturate the readout link; when this happens, the acquisition is paused causing dead time. Real and Dead time counters are available.\\

\begin{figure}
	\begin{center}
	\includegraphics[width=7.0in,angle=90]{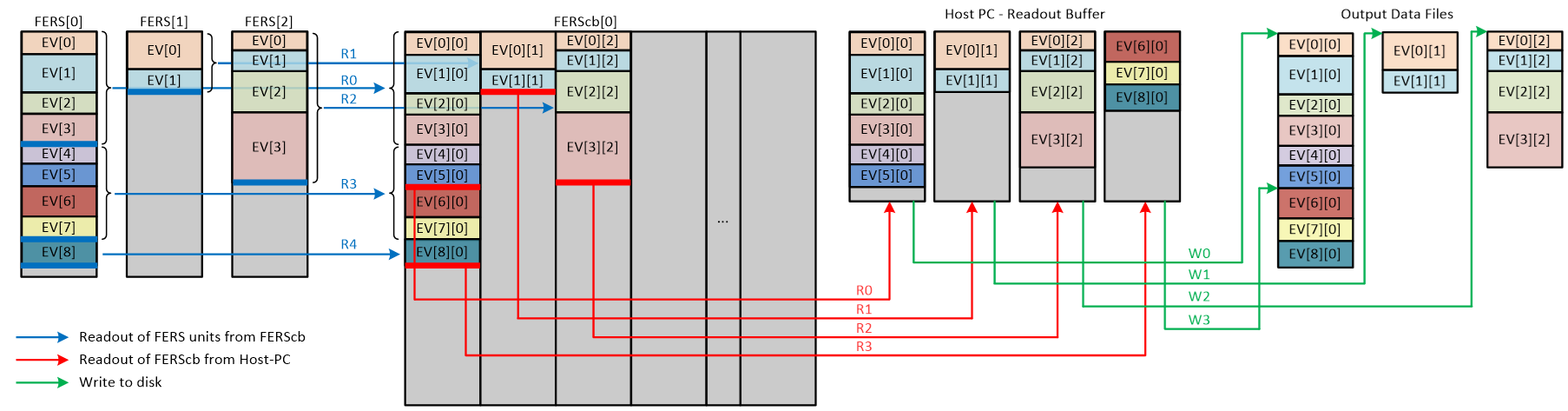}
	\end{center}
	\caption{Example of Vertical Readout with 3 FERS acquiring at different rates (0 high, 1 and 2 low)}
	\label{vertical_readout}
\end{figure}

Since there is not a global trigger and the data throughput of one FERS unit can be significantly different from another one, the data readout takes place “vertically” (see Fig. \ref{vertical_readout}): the software reads a block of data from each unit, but there is not any correlation between the data coming from different units. The software gives priority to the units that are producing more data; silent units will be skipped until they have data to read.\\

\subsubsection{Timing and ToT mode}
This mode is similar to the Timing mode, with added energy information obtained through Time-Over-Threshold (ToT) for low resolution spectroscopy. ToT and individual time stamps are saved for the channels that have been fired. For each trigger, the FPGA opens an acquisition window of programmable size and acquires the time stamps of the self-triggers occurring within the window. The time stamp has a resolution of 0.5 ns. The channels that are not fired are suppressed and give no data. Also in this mode, the output data is a list of time-stamped events grouped in packets of programmable size. The Timing and ToT mode supports the same 3 sub-mode implemented in the Timing Mode. Besides the time stamp, the output data contains the ToT value. From these values a low-resolution energy spectrum can be reconstructed. In contrast with Spectroscopy mode, it is possible to sustain higher event rates (max total rate $\sim$ 20 Mcps) because the A/D conversion of the input signal is not done.

\section{Preliminary results}
In the following, the first preliminary results obtained measuring the output of a Hamamatsu S13361 series 64-channel SiPM matrix using the A5202 running in Spectroscopy and ToT mode is shown.\\
The SiPM matrix was connected to the A5202 using the A5251 MPPC adapter compatible with Hamamatsu pinout. The matrix was biased at about 55 V.

\begin{figure}
	\begin{center}
	\includegraphics[width=3.5in]{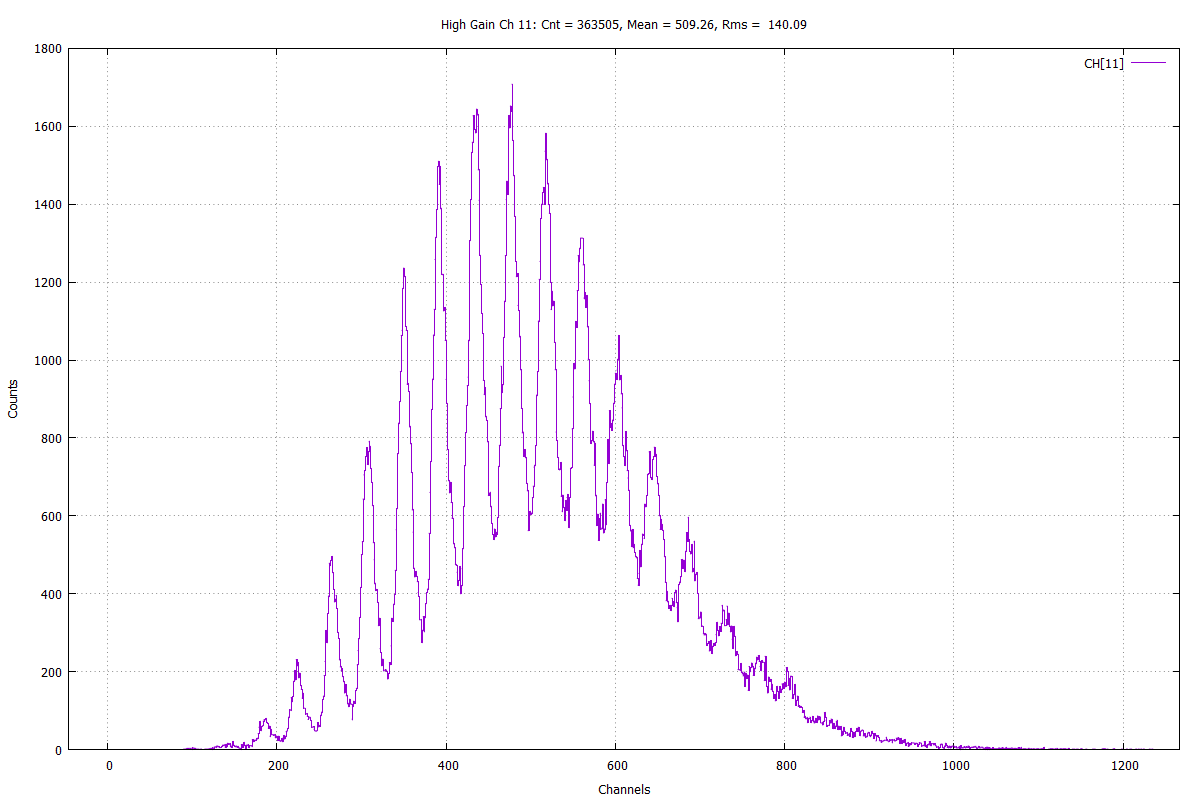}
	\end{center}
	\caption{Multiphoton spectrum from one channel of the A5202, obtained by illuminating Hamamatsu S13361 matrix with CAEN SP5601 LED driver. The A5202 was running in Spectroscopy mode and the spectrum was acquired with FERS Readout Software. Multiphoton peaks are clearly visible.}
	\label{Multiphoton_peak_pha}
\end{figure}

\begin{figure}
	\begin{center}
	\includegraphics[width=3.5in]{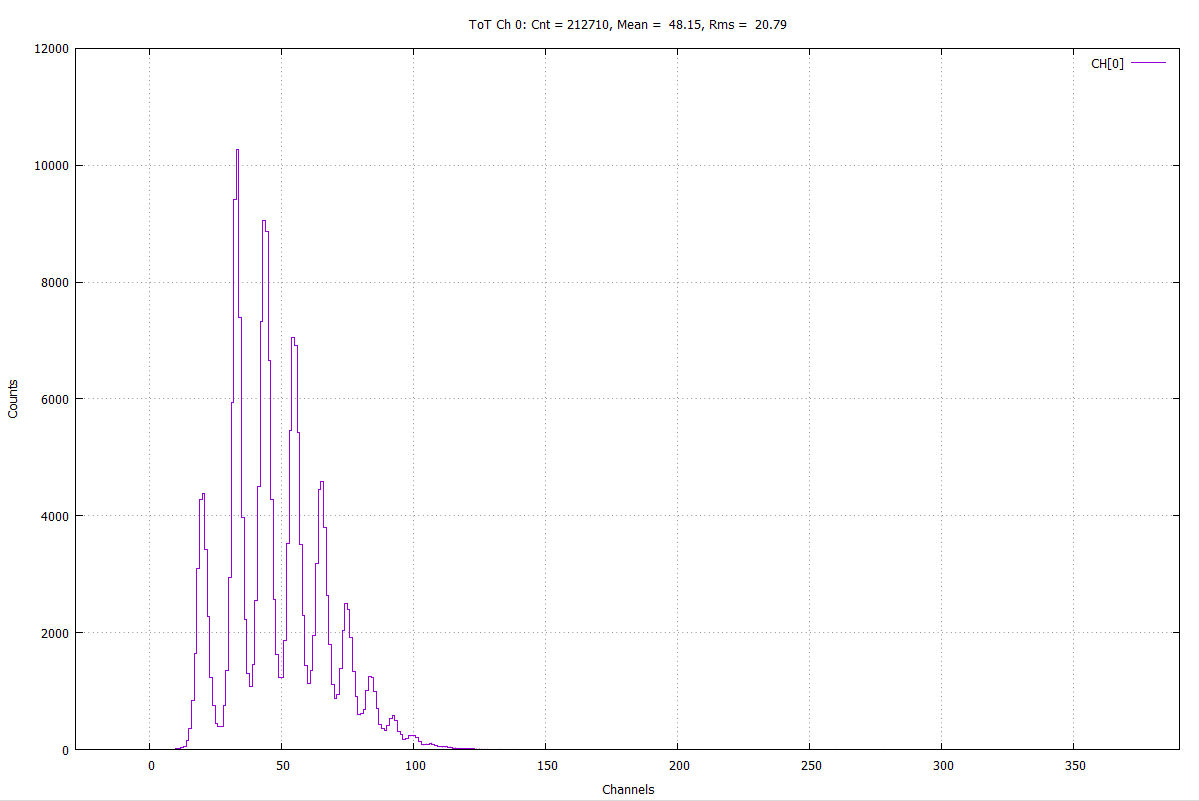}
	\end{center}
	\caption{Energy spectrum from one high-gain channel of the A5202, obtained by illuminating Hamamatsu S13361 matrix with CAEN SP5601 LED driver. The A5202 was running in ToT mode and the spectrum was acquired with FERS Readout Software. Multiphoton peaks are clearly visible.}
	\label{ToT}
\end{figure}

Fig.\ref{Multiphoton_peak_pha} and \ref{ToT} show multiphoton spectrum from one channel of the A5202 obtained by illuminating the SiPM matrix with CAEN SP5601 LED driver\cite{SP5601} and get in Spectroscopy and ToT mode respectively.

\begin{figure}
	\begin{center}
	\includegraphics[width=3.5in]{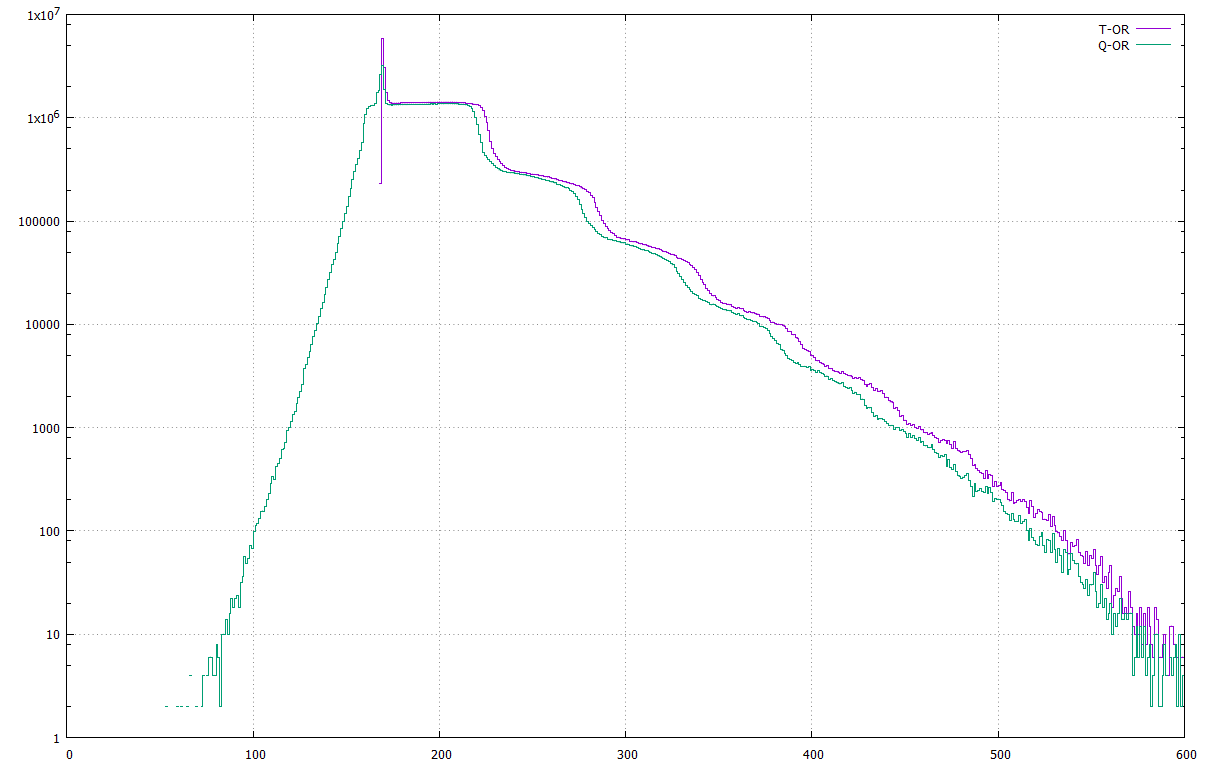}
	\end{center}
	\caption{Counts vs Threshold (Staircase) from the OR of the SiPM matrix pixels connected to one high-gain channel of the A5202.}
	\label{1_scan_thr_trg_off_pa_0}
\end{figure}

Fig.\ref{1_scan_thr_trg_off_pa_0} shows the Counts vs Threshold (Staircase) plot get from the OR of the SiPM matrix pixels connected to one high-gain channel of the A5202, obtained by illuminating Hamamatsu S13361 matrix with CAEN SP5601 LED driver. The two curves are relative at the threshold set on the two available discriminators: fast (T-OR, magenta) and slow (Q-OR, green). It worth to notice the different behaviors of the two staircase: the Q-OR one goes below minimum threshold of the T-OR one. This is due to the slow discriminator filter signal shape that shows a sort of undershoot on which make the trigger fires. This undershoot is not present in the T-OR fast discriminator signal shape.

Furthermore, a second set of preliminary measurement have been performed using an external pulser. They were about the A5202 ADC and ToT Non-Linearity. Each channel of Citiroc 1A embeds two channel-by-channel independent programmable variable-gain preamplifiers ensuring a wide coverage of the dynamic range depending on the application. Both low gain and high gain preamplifier have been tested. \\ 
The test condition are here summarized:

\begin{enumerate}
	
	\item Low Gain mode
	\begin{itemize}
	\item Input Signal:
		\begin{itemize}
		\item Shape: square wave for different rise and fall time
		\item Frequency = 100 Hz
		\item Rise Time = 10 ns
		\item Fall Time = 100 $\mu$s
		\end{itemize}
	\item Variable amplitude ($V_{in}$) from 100 mV to 4900 mV in steps of 300 mV
	\item Input capacity = 100 pF
	\item 1 pe = 10$^6$ electrons = 160 fC (assuming SiPM Gain = 10$^6$)
	\item Citiroc Configuration:
		\begin{itemize}  
		\item Shaping Time = 0 (87.5ns)
		\item Fast discriminator T Coarse Threshold = 175 
		\item Fast discriminator T Fine Threshold = 0
		\item Low Gain = 36
		\item Trigger Hold Time = 100 ns
		\end{itemize}
	\end{itemize}

	\item High Gain mode
	\begin{itemize}
		\item Input Signal:
		\begin{itemize}
			\item Shape: square wave for different rise and fall time
			\item Frequency = 100 Hz
			\item Rise Time = 10 ns
			\item Fall Time = 100 $\mu$s
		\end{itemize}
		\item Variable amplitude ($V_{in}$) from 10mV to 120mV in steps of 10mV
		\item Input capacity = 100 pF
		\item 1 pe = 10$^6$ electrons = 160 fC (assuming SiPM Gain = 10$^6$)
		\item Citiroc Configuration:
		\begin{itemize}  
			\item Shaping Time = 5 (25ns)
			\item Fast discriminator T Coarse Threshold = 185 
			\item Fast discriminator T Fine Threshold = 0
			\item Low Gain = 51
			\item Trigger Hold Time = 200 ns	
		\end{itemize}
	\end{itemize}

\end{enumerate}

\begin{figure}
	\begin{center}
		\includegraphics[width=3.0in]{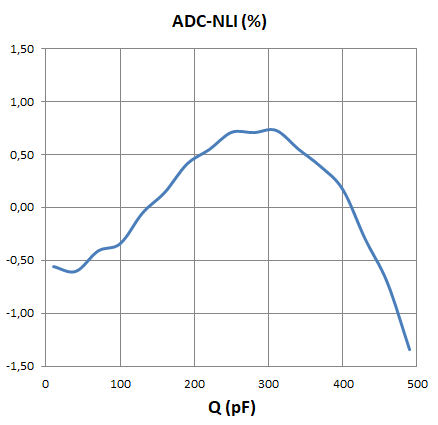}
	\end{center}
	\caption{A5202 ADC Non-Linearity in Low Gain mode.}
	\label{ADC_NLI}
\end{figure}

\begin{figure}
	\begin{center}
	\includegraphics[width=3.0in]{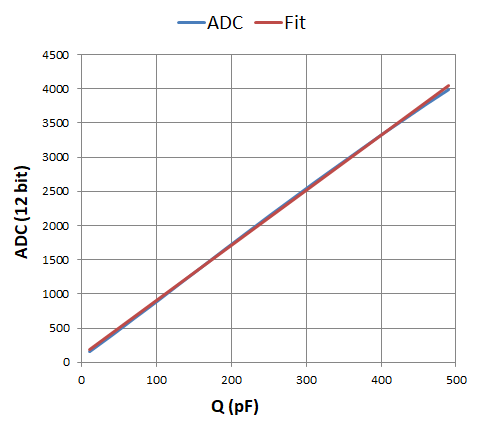}
	\end{center}
	\caption{Comparison between the A5202 ADC mean value and the corresponding value of the linear function that fits the ADC values in Low Gain mode.}
	\label{ADC_Fit}
\end{figure}

\begin{figure}
	\begin{center}
	\includegraphics[width=3.0in]{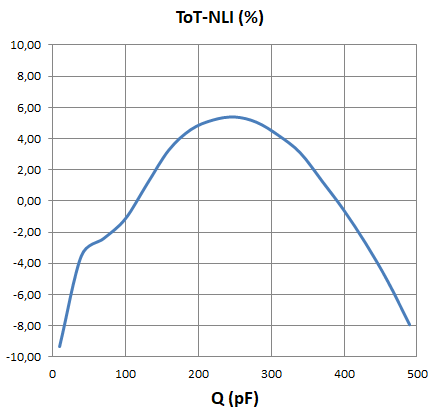}
	\end{center}
	\caption{A5202 non linearity when running in ToT mode.}
	\label{ToT_NLI}
\end{figure}

\begin{figure}
	\begin{center}
	\includegraphics[width=3.0in]{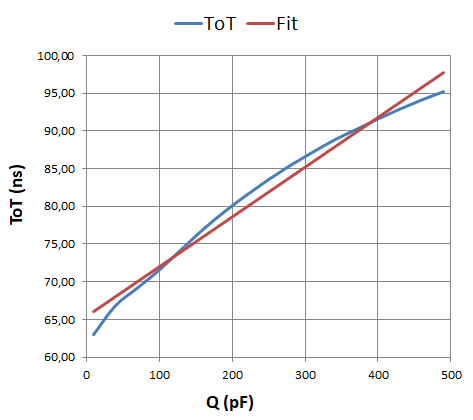}
	\end{center}
	\caption{Comparison between the A5202 ToT mean value and the corresponding value of the linear function that fits the ToT values.}
	\label{ToT_Fit}
\end{figure}

\begin{figure}
	\begin{center}
		\includegraphics[width=3.0in]{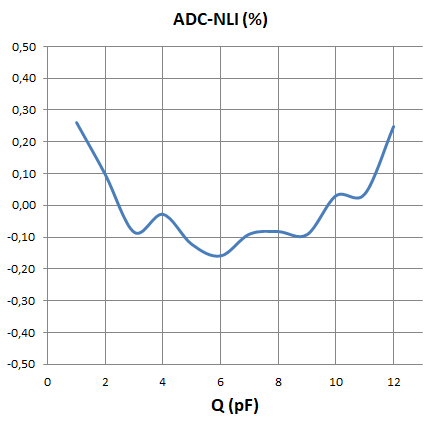}
	\end{center}
	\caption{A5202 ADC Non-Linearity in Low Gain mode.}
	\label{ADC_NLI_HighGain}
\end{figure}

\begin{figure}
	\begin{center}
		\includegraphics[width=3.0in]{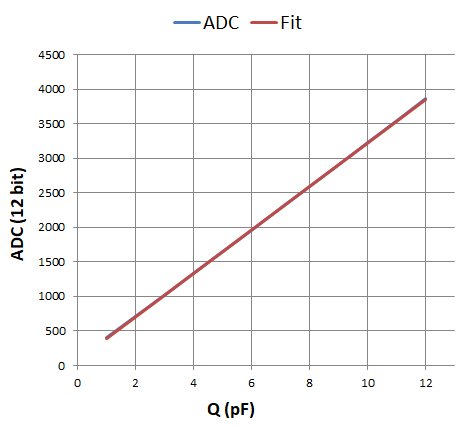}
	\end{center}
	\caption{Comparison between the A5202 ADC mean value and the corresponding value of the linear function that fits the ADC values in Low Gain mode.}
	\label{ADC_Fit_HighGain}
\end{figure}

The results are shown in Fig.\ref{ADC_NLI}, \ref{ADC_Fit}, \ref{ToT_NLI} and \ref{ToT_Fit} for the Low Gain mode and in Fig.\ref{ADC_NLI_HighGain}, \ref{ADC_Fit_HighGain} for the High Gain case. The latter has still to be completed with the ToT Non-Linearity evaluation.\\ 

Fig.\ref{ADC_NLI}, \ref{ADC_NLI_HighGain} and \ref{ToT_NLI} show the ADC (in Low Gain and High Gain mode respectively) and the TOT Non-Linearity while Fig. \ref{ADC_Fit} and \ref{ToT_Fit} show, in the Low Gain mode, per each value of the injected charge, the comparison between the average ADC measured values and the value of the linear function used to fit them.

The ADC Non-Linearity is within 0.5$\%$ and 1.5$\%$ both in Low and High Gain mode; the ToT Non-Linearity is within 10$\%$ in Low Gain mode while the High Gain mode is still to be characterized.

The ADC Non-Linearity linear fit RMS is, expressed in photoelectron (p.e.), about 6,8 and 0,34 in the Low Gain and in the High Gain mode respectively; the ToT Non-Linearity linear fit RMS is about 47,2 p.e. in Low Gain mode while the High Gain mode is still to be characterized. 

These results show that only in High Gain mode it is possible to go down to the single photoelectron sensitivity even though some optimization on the A5202 noise level has been done and a new more extensive characterization campaign will be done in order to better evaluate the platform performance.

\section{Conclusions}
FERS-5200 is the result of the effort to provide, not a simple readout board, but a platform to allow for the readout of thousand of channels with easy multiboard synchronization, reduced cost per channel and a straightforward passage from a small evaluation system to a huge experimental installation. The different readout modes allow to explore several applications, maintaining the same hardware setup and DAQ.\\ Thanks to the hardware architecture of the FERS units, further developments will be possible by simply replacing new ASICs or using an hybrid solution with Front-End in a separate box and plug-in FERS with Flash ADC. In this way, we aim to rapidly have a wide variety of boards for different detectors and applications.

\end{document}